\documentstyle[prb,aps,psfig,floats]{revtex}
\begin{document}
\twocolumn
\draft
\wideabs{
\title
{c-axis penetration depth in Bi$_2$Sr$_2$CaCu$_2$O$_{8+\delta}$
single crystals measured by ac-susceptibility and cavity perturbation
technique}

\author{D.~V.~Shovkun, M.~R.~Trunin, A.~A.~Zhukov, and Yu.~A.~Nefyodov}
\address{Institute of Solid State Physics, 142432, Chernogolovka,
Moscow district, Russia}

\author{N.~Bontemps and H.~Enriquez}
\address{Laboratoire de Physique de la Mati\`ere Condens\'ee, Ecole Normale
Sup\'erieure, 24 rue Lhomond, 75231 Paris Cedex 05, France}

\author{A.~Buzdin and M.~Daumens}
\address{Laboratoire de Physique Th\'eorique, Universit\'e Bordeaux I,
33405 Talence Cedex, France}

\author{T.~Tamegai}
\address{Department of Applied Physics, The University of Tokyo,
Hongo, Bunkyo-ku, 113-8656, Japan}

\date\today
\maketitle
\begin{abstract}
The $c$-axis penetration depth
$\Delta\lambda_c$ in Bi$_2$Sr$_2$CaCu$_2$O$_{8+\delta}$ (BSCCO) single
crystals as a function of temperature has been determined using two
techniques, namely, measurements of the ac-susceptibility at a frequency
of 100~kHz and the surface impedance at 9.4~GHz. Both techniques yield
an almost linear function $\Delta\lambda_c(T)\propto T$ in the
temperature range $T<0.5\,T_c$. Electrodynamic analysis of the impedance
anisotropy has allowed us to estimate $\lambda_c(0)\approx 50$~$\mu$m in
BSCCO crystals overdoped with oxygen ($T_c\approx 84$~K) and
$\lambda_c(0)\approx 150$~$\mu$m at the optimal doping level
($T_c\approx 90$~K).

\end{abstract}
}

Cuprate high-temperature superconductors (HTS) are layered anisotropic
materials. Therefore the electrodynamic problem of the magnetic field
penetration depth in HTS in the low-field limit is characterized by two
length parameters, namely, $\lambda_{ab}$ controlled by screening
currents running in the CuO$_2$ planes (in-plane penetration depth)
and $\lambda_c$ due to currents running in the direction perpendicular
to these planes (out-of-plane or $c$-axis penetration depth). The
temperature dependence of the penetration depth in HTS is largely
determined by the superconductivity mechanism. It is known
(see, e.g., Ref.~\cite{Tru1} and references therein) that
$\Delta\lambda_{ab}(T)\propto T$ in the range $T<T_c/3$ in
high-quality HTS samples at the optimal level of doping, and this
observation has found the most simple interpretation in the $d$-wave
model of the high-frequency response in HTS \cite{Scal}.
Measurements of $\lambda_c(T)$ are quoted less frequently than
those of $\lambda_{ab}(T)$. Most of such data published by far
were derived from microwave measurements of the surface impedance of HTS
crystals \cite{Shib1,Mao,Kit1,Bon1,Jac1,Shib2,Srik,Kit2,Hos}. There is
no consensus in literature about $\Delta\lambda_c(T)$ at low
temperatures. Even in reports on low-temperature properties of
high-quality YBCO crystals, which are the most studied objects, one can
find both linear, $\Delta\lambda_c(T)\propto T$ \cite{Mao,Srik}, and
quadratic dependences \cite{Hos} in the range $T<T_c/3$. In BSCCO
materials, the shape of $\Delta\lambda_c(T)$ depends on the level of
oxygen doping: in samples with maximal $T_c\simeq 90$~K
$\Delta\lambda_c(T)\propto T$ at low temperatures \cite{Jac1,Shib2}; at
higher oxygen contents (overdoped samples) $T_c$ is lower and the
linear function $\Delta\lambda_c(T)$ transforms to a quadratic one
\cite{Shib2}. The common feature of all microwave experiments is that
the change in the ratio $\Delta\lambda_c(T)/\lambda_c(0)$ is smaller
than in $\Delta\lambda_{ab}(T)/\lambda_{ab}(0)$ because in all HTS
$\lambda_c(0)\gg\lambda_{ab}(0)$.
The length $\lambda_c(0)$ is especially large in BSCCO crystals,
$\lambda_c(0)>10$~$\mu$m and, according to some estimates, it ranges up
to $\sim 500$~$\mu$m. The large spread of $\lambda_c(0)$ is caused by
two factors, namely, the poor accuracy of the techniques used in
determination of $\lambda_c(0)$ and effects of local and extended
defects in tested samples, whose range is of order of 1~mm and
comparable to both $\lambda_c$ and total sample dimensions.

Recently we suggested \cite{Nic} a new technique for determination
of $\lambda_c(0)$ based on the measurements of the surface
barrier field $H_J(T)\propto 1/\lambda_c(T)$ at which
Josephson vortices penetrate into the sample. The field $H_J$
corresponds to the onset of microwave absorption in the locked state of
BSCCO single crystals. This paper suggests an alternative technique
based on comparison between microwave measurements of BSCCO crystals
aligned differently with respect to ac magnetic field and a numerical
solution of the electrodynamic problem of the magnetic field distribution
in an anisotropic plate at an arbitrary temperature. Moreover, since
$\lambda_c(0)$ in BSCCO single crystals is relatively large, we managed
to determine $\lambda_c(T)$ from the temperature dependences of
ac-susceptibility and compare these measurements to results of
microwave experiments.

Single crystals of BSCCO were grown by the floating-zone method
\cite{Tam} and shaped as rectangular platelets. This paper presents
measurements of two BSCCO samples with various levels of oxygen doping.
The first sample (\#1), characterized by a higher critical temperature,
$T_c\approx 90$~K (optimally doped), has dimensions
$a\times b\times c\simeq 1.5\times 1.5\times 0.1$~mm$^3$ ($a\approx b$).
The second (\#2, $a\times b\times c\simeq 0.8\times 1.8\times 0.03$~mm$^3$)
is slightly overdoped ($T_c\approx 84$~K).

When measuring the ac-susceptibility $\chi=\chi'-i\chi''$, we placed
a sample inside one of two identical induction coils. The coils were
connected to one another, and the out-of-phase and in-phase components
of the imbalance signal were measured at a frequency of $10^5$~Hz.
These components are proportional to the real and
imaginary parts of the sample magnetic moment $M$, respectively:
$M=\chi vH_0$, where $v$ is the sample volume and $H_0$ is the ac
magnetic field amplitude, which was within 0.1~Oe in our experiments.

Figure~1 shows temperature dependences $\chi'(T)/|\chi'(0)|$ in
\begin{figure}[h]
\centerline{\psfig{figure=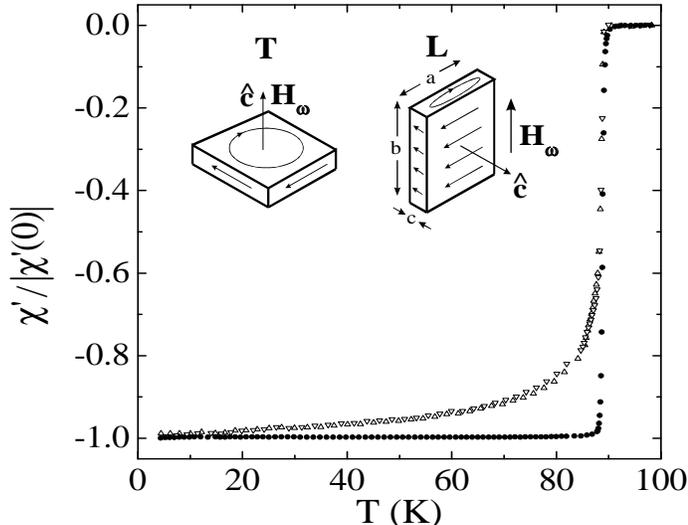,height=7cm,width=9cm,clip=,angle=0.}}
\caption{Curves of the ac-susceptibility of sample~\#1 versus
temperature in different orientation with respect to ac magnetic field:
${\bf H}_{\omega}\parallel {\bf c}$ (full circles);
${\bf H}_{\omega}\perp {\bf c}$, ${\bf H}_{\omega}$ is parallel to
the $b$-edge of the crystal (up triangles);
${\bf H}_{\omega}\perp {\bf c}$, ${\bf H}_{\omega}$ is
parallel to the $a$-edge of the crystal (down triangles).
Left-hand inset: transverse (T) orientation,
${\bf H}_{\omega}\parallel {\bf c}$, the arrows on the surfaces show
directions of the screening current. Right-hand inset: Longitudinal (L)
orientation, ${\bf H}_{\omega}\perp {\bf c}$.}
\label{f1}
\end{figure}
sample~\#1 for three different sample alignments with respect to
the ac magnetic field: the transverse (T) orientation,
${\bf H}_{\omega}\parallel {\bf c}$, (the inset on the left of Fig.~1),
when the screening current flows in the $ab$-plane (full circles); in the
longitudinal (L) orientation, ${\bf H}_{\omega}\perp {\bf c}$, (the
inset on the right of Fig.~1, ${\bf H}_{\omega}$ is parallel to the
$b$-edge of the crystal), when currents running in the directions of
both CuO$_2$ planes and the $c$-axis are present (up triangles); in the
L-orientation, ${\bf H}_{\omega}\perp {\bf c}$, whose difference from
the previous configuration is that the sample is turned around the
$c$-axis through $90^\circ$ (down triangles). Fig.~1 clearly shows that at
$T<T_c$ $\chi'_{ab}(T)$ is notably smaller in the T-orientation than
$\chi'_{ab+c}(T)$ in the L-orientation (the subscripts of $\chi'$ denote
the direction of the screening current). The coincidence
of $\chi'_{ab+c}(T)$ curves at ${\bf H}_{\omega}\perp {\bf c}$
and the small width of the superconducting transition at
${\bf H}_{\omega}\parallel {\bf c}$
($\Delta T_c< 1$~K) indicate that the quality of the tested
sample~\#1 is fairly high. This is supported by precision measurements
of surface impedance $Z_s(T)=R_s(T)+iX_s(T)$ of sample~\#1 at frequency
$f=9.4$~GHz in the T-orientation, which are plotted in Fig.~2. The
measurement technique was described in detail elsewhere \cite{Tru1}.
It applies to both surface impedance components $R_s(T)$ and $X_s(T)$:
\begin{figure}[t]
\centerline{\psfig{figure=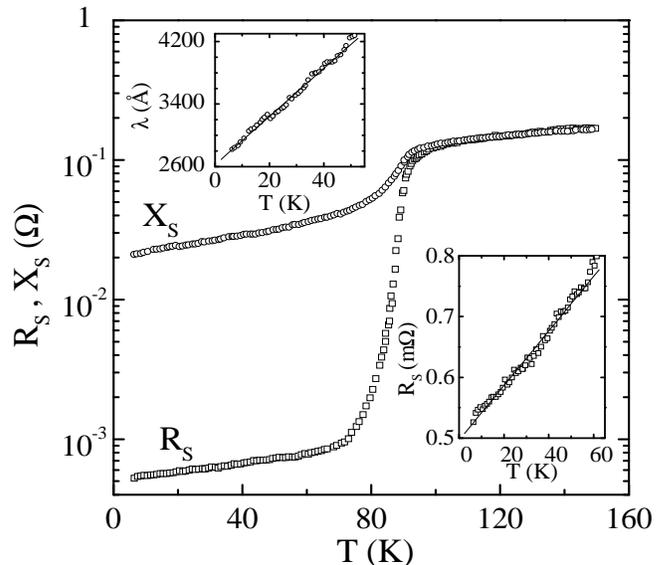,height=7.5cm,width=8.5cm,clip=,angle=0.}}
\caption{Surface resistance $R_s(T)$ and reactance $X_s(T)$ in $ab$-plane
(T-orientation) of sample~\#1 at $9.4$ GHz. The insets show
linear plots of $\lambda(T)$ and $R_s(T)$ at low temperatures. }
\label{f2}
\end{figure}

\begin{equation}
R_s=\Gamma_s\,\Delta(1/Q),\qquad X_s=-2\,\Gamma_s\,\delta f/f,\label{RX}
\end{equation}
Here $\Gamma_s=\omega \mu_0\int_V H^2_{\omega}dV /[\int_S
H_s^2 dS]$ is the sample geometrical factor ($\omega=2\pi f$,
$\mu_0=4\pi\cdot10^{-7}$~H/m, $V$ is the volume of the cavity,
$H_{\omega}$ is the magnetic field generated in the
cavity, $S$ is the total sample surface area, and $H_s$ is the
tangential component of the microwave magnetic field on the sample
surface); $\Delta (1/Q)$ is the difference between the values
1/Q of the cavity with the sample inside and empty cavity; $\delta f$
is the frequency shift relative to that which would be measured for a
sample with perfect screening, i.e., no penetration of the microwave
fields. In the experiment we measure the difference $\Delta f(T)$
between resonant frequency shifts with temperature of the loaded and
empty cavity, which is equal to $\Delta f(T)=\delta f(T)+f_0$, where
$f_0$ is a constant \cite{exp}. The constant $f_0$ includes both
the perfect-conductor shift and the uncontrolled contribution caused by
opening and closing the cavity. In HTS single crystals, the constant
$f_0$ can be directly derived from measurements of the surface impedance
in the normal state; in particular, in the T-orientation $f_0$ can be
derived from the condition that the real and imaginary parts of the
impedance should be equal above $T_c$ (normal skin-effect).
In Fig.~2 $R_s(T)=X_s(T)$ at $T\ge T_c$, and its temperature dependence
is adequately described by the expression $2R_s^2(T)/\omega
\mu_0=\rho(T)=\rho_0+bT$ with $\rho_0\approx 13$~$\mu\Omega\cdot$cm and
$b\approx 0.3$~$\mu\Omega\cdot$cm/K. Given $R_s(T_{\rm c})=\sqrt{\omega
\mu_0 \rho(T_{\rm c})/2}\approx 0.12$~$\Omega$, we obtain the
resistivity $\rho(T_{\rm c})\approx 40$~$\mu\Omega\cdot$cm. The insets
to Fig.~2 show $R_s(T)$ and $\lambda (T)=X_s(T)/\omega\mu_0$ for
$T<0.7\,T_{\rm c}$ plotted on a linear scale. The extrapolation of the
low-temperature sections of these curves to $T=0$ yields estimates of
$\lambda_{ab}(0)\approx 2600$~\AA~and the residual surface resistance
$R_{\rm res}\approx 0.5$~m$\Omega$. $R_{\rm res}$ is due to various defects
in the surface layer of the superconductor and it is generally accepted that
the lower the $R_{{\rm res}}$, the better the sample quality.
The above mentioned parameters of sample~\#1 indicate that its quality
is fairly high. In the T-orientation, linear functions $\Delta R_s(T)$
and $\Delta\lambda_{ab}(T)$ in the low-temperature range were previously
observed in optimally doped BSCCO crystals at a frequency of about
10~GHz \cite{Jac1,Shib2,Lee}. In the slightly overdoped sample~\#2 we
also observed $\Delta\lambda_{ab}(T)$, $\Delta R_s(T)\propto T$ at low
temperature, moreover, the measurement
$R_{\rm {res}}\approx 120$~$\mu\Omega$ is, to the best of our knowledge,
the lowest value ever obtained in BSCCO single crystals.

In both superconducting and normal states of HTS, the relation between
electric field and current density is local: $j=\hat\sigma E$, where the
conductivity $\hat\sigma$ is a tensor characterized by components
$\sigma_{ab}$ and $\sigma_c$. In the normal state, ac field penetrates
in the direction of the $c$-axis through the skin depth
$\delta_{ab}=\sqrt{2/\omega\mu_0\sigma_{ab}}$ and in the CuO$_2$ plane
through $\delta_c=\sqrt{2/\omega\mu_0\sigma_c}$. In the
superconducting state all parameters $\delta_{ab}$, $\delta_c$,
$\sigma_{ab}=\sigma'_{ab}-i\sigma''_{ab}$, and $\sigma_c=\sigma'_c
-i\sigma''_c$ are complex. In the
temperature range $T<T_c$, if $\sigma'\ll \sigma''$, the field
penetration depths are given by the formulas
$\lambda_{ab}=\sqrt{1/\omega\mu_0\sigma''_{ab}}$,
$\lambda_c=\sqrt{1/\omega\mu_0\sigma''_c}$. In the close neighborhood of
$T_c$, if $\sigma'\agt \sigma''$, the decay of magnetic field in
the superconductor is characterized by the functions
$\rm {Re}\,(\delta_{ab})$ and $\rm {Re}\,(\delta_c)$, which turn to
$\delta_{ab}$ and $\delta_c$, respectively, at $T\ge T_c$.

In the L-orientation of BSCCO single crystals at $T<0.9\,T_c$ the
penetration depth is smaller than characteristic sample dimensions.
If we neglect the anisotropy in the $ab$-plane and the contribution
from $ac$-faces (see the inset to Fig.~1), which is a factor $\sim c/b$
smaller than that of the $ab$-surfaces, the effective impedance
$Z_s^{ab+c}$ in the L-orientation can
be expressed in terms of $Z_s^{ab}$ and $Z_s^c$ averaged over the
surface area \cite{Tru1,Kit1} (the superscripts of $Z_s$ denote the
direction of the screening current). Thus, given measurements of
$\Delta\lambda_{ab}(T)=\Delta X_s^{ab}(T)/\omega\mu_0$ in the
T-orientation and of the effective value $\Delta\lambda_{ab+c}(T)=\Delta
X_s^{ab+c}(T)/\omega\mu_0$ in the L-orientation, we obtain
\begin{equation}
\Delta\lambda_c=\left[(a+c)\,\Delta\lambda_{ab+c}-
a\,\Delta\lambda_{ab}\right]/c~.
\label{DL}
\end{equation}

This technique for determination of $\Delta\lambda_c(T)$ was used in
microwave experiments \cite{Shib1,Mao,Kit1,Bon1,Jac1,Shib2,Srik} at low
temperatures, $T<T_c$. Even so, it cannot be applied to the range
of higher temperatures because the size effect plays an important role.
Really, at $T>0.9\,T_{\rm c}$ the lengths $\lambda_c$ and $\delta_c$ are
comparable to the sample dimensions. In order to analyze our
measurements in both superconducting and normal states of BSCCO, we used
formulas \cite{Gou} for field distributions in an anisotropic long strip
($b\gg{a,c}$) in the L-orientation. These formulas neglect the effect of
$ac$-faces of the crystal, but take account of the size effect.
In addition, in a sample shaped as a long strip, there is a
simple relation between its surface impedance components and
ac-susceptibility, which is expressed in terms of parameter $\mu$
introduced in Ref.~\cite{Gou}:
\begin{equation}
\Delta\,(1/Q)-2i\,\delta f/f=i \gamma \mu v/V,\qquad \chi=-1+\mu,
\label{QF}
\end{equation}
where $\gamma=VH_0^2/[\int_V H^2_{\omega}dV]=10.6$ is a constant
characterizing our cavity \cite{Tru1}. At an arbitrary
temperature, the complex parameter $\mu=\mu'-i\mu''$ is
controlled by the components $\sigma_{ab}(T)$ and $\sigma_c(T)$
of the conductivity tensor:
\begin{equation}
\mu={8 \over \pi^2}\sum_n {1\over n^2}\left\{
{\tan(\alpha_n) \over \alpha_n}+
{\tan(\beta_n) \over \beta_n}
\right\},
\label{MU}
\end{equation}
where the sum is performed over odd integers $n>0$, and
\begin{eqnarray}
\alpha_n^2=-{a^2 \over \delta_c^2}
\left({i\over 2}+{\pi^2 \over 4}{\delta_{ab}^2 \over c^2}n^2 \right),\,
\beta_n^2=-{c^2 \over \delta_{ab}^2}
\left({i\over 2}+{\pi^2 \over 4}{\delta_c^2 \over a^2}n^2 \right).
\nonumber
\end{eqnarray}

In the superconducting state at $T<0.9\,T_c$ we find that
$\lambda_{ab}\ll c$ and $\lambda_c\ll a$.
In this case, we derive from Eq.~(\ref{MU}) a simple
expression for the real part of $\mu$:
\begin{equation}
\mu'=1+\chi'={2\lambda_c \over a}+ {2\lambda_{ab} \over c}~.
\label{CH}
\end{equation}

One can easily check up that in the range of low temperatures the change
in $\Delta\lambda_{\rm c}(T)$ prescribed by Eq.~(5) is identical to
Eq.~(2). Figure~3 shows measurements of $\Delta\lambda_c(T)$ in
sample~\#1 (circles) and sample~\#2 (squares) at $T<0.9\,T_c$. The
\begin{figure}[h]
\centerline{\psfig{figure=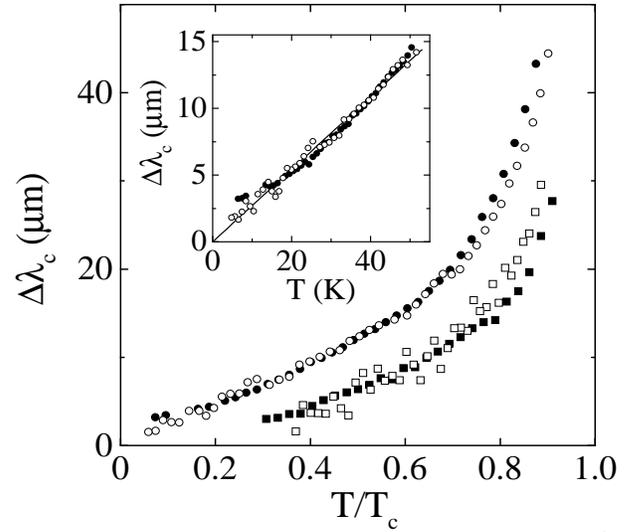,height=7cm,width=8cm,clip=,angle=0.}}
\caption{Temperature dependences $\Delta\lambda_c$ in samples~\#1
(circles) and \#2 (squares) at $T<0.9\,T_c$. Open symbols plot
low-frequency measurements, full symbols show microwave data.
The inset shows low temperature sections of the $\Delta\lambda_c$ curves
in sample~\#1.}
\label{f3}
\end{figure}
open symbols plot low-frequency measurements obtained in accordance with
Eq.~(5), the full symbols plot microwave measurements processed by
Eq.~(2). Agreement between measurements of sample~\#2 (lower curve) is
fairly good, but in fitting together experimental data from sample~\#1
(upper curve) we had to divide by a factor of 1.8 all
$\Delta\lambda_c(T)$ derived from measurements of ac-susceptibility
using Eq.~(5). The cause of the difference between $\Delta\lambda_c(T)$
measured in sample~\#1 at different frequencies is not quite clear. We
rule out a systematic experimental error that could be caused by
misalignment of the sample with respect to the ac magnetic field in the
coil because (i) curves of $\Delta\lambda_c(T)$ were accurately
reproduced when square sample~\#1 was turned through an angle of
$90^\circ$ in the L-orientation, and (ii) a small tilt of this sample
with respect to the magnetic field generated by the coil would lead to a
larger difference (more than a factor of 1.8) between the two sets of
experimental data. It seems more plausible that Eqs.~(2) and (5), which
neglect the contribution of $ac$-faces, yield inaccurate results
concerning sample~\#1: its $ac$-faces, which have a notable area
(sample~\#1 is thick), can host a lot of defects (for example, those of
the capacitive type), and the latter can affect the character of field
penetration as a function of frequency.

The curves of $\Delta\lambda_c(T)$ at $T<0.5\,T_c$ plotted in Fig.~3
are almost linear: $\Delta\lambda_c(T)\propto T$. The inset to Fig.~3
shows the low-temperature section of the curve of $\Delta\lambda_c(T)$
in sample~\#1. Its slope is 0.3~$\mu$m/K and equals that from
Ref.~\cite{Shib2}. Note also that changes in
$\Delta\lambda_c(T)$ are smaller in the oxygen-overdoped sample~\#2 than
in sample~\#1.

We can estimate $\lambda_c(0)$ by substituting in Eq.~(1) $\delta f(0)$
obtained by comparing of $\Delta(1/Q)$ and $\Delta f=\delta f-f_0$
measurements taken in the T- and L-orientations to numerical calculations
by Eqs.~(3) and (4), which take account of the size effect in
the high-frequency response of an anisotropic crystal. The procedure of
comparison for sample~\#1 is illustrated by Fig.~4. Unlike the case of
the T-orientation, the measured temperature dependence of $\Delta(1/Q)$
\begin{figure}[h]
\centerline{\psfig{figure=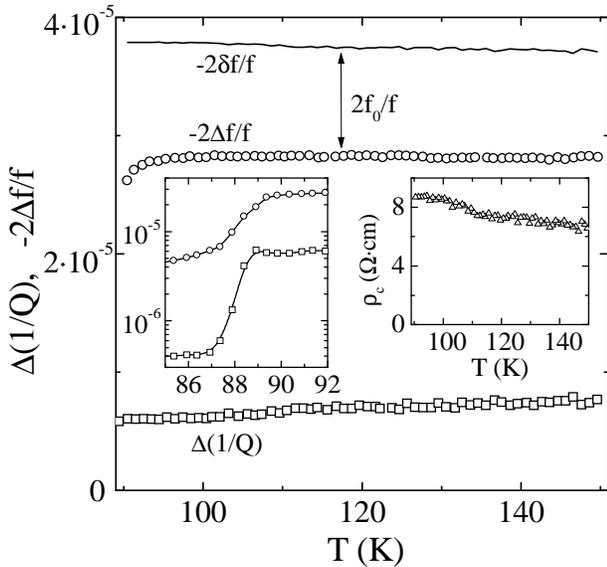,height=7.5cm,width=8cm,clip=,angle=0.}}
\caption{Temperature dependences of $\Delta (1/Q)$ (open squares) and
$-2\Delta f/f$ (open circles) in the L-orientation of sample~\#1 at
$T>T_c$. Solid line shows the function $-2\delta f(T)/f$ deriving from
Eq.~(\protect\ref{QF}). Left-hand inset: $\Delta
(1/Q)$ and $-2\Delta f/f$ as functions of temperature in the
neighborhood of $T_c$. Right-hand inset: $\rho_c(T)$ in sample~\#1
(triangles).}
\label{f4}
\end{figure}
in the L-orientation deviates from $(-2\Delta f/f)$ owing to the size
effect. Using the measurements of
$R_s=\sqrt{\omega\mu_0/2\sigma_{ab}}$ at $T>T_c$ in the
T-orientation (Fig.~2) for determination of $\sigma_{ab}(T)$, alongside
the data on $\Delta(1/Q)$ in the L-orientation (open squares in Fig.~4),
from Eqs.~(3) and (4) we obtain the curve of $\rho_c(T)=1/\sigma_c(T)$
shown in the right-hand inset to Fig.~4. Further, using the functions
$\sigma_c(T)$ and $\sigma_{ab}(T)$, we calculate $(-2\delta f/f)$ versus
temperature for $T>T_c$, which is plotted by the solid line in Fig.~4.
This line is approximately parallel to the experimental curve of
$-2\Delta f/f$ in the L-orientation (open circles in Fig.~4). The
difference $-2(\delta f-\Delta f)/f$ yields the additive constant $f_0$.
Given $f_0$ and $\Delta f(T)$ measured in the range $T<T_c$, we also obtain
$\delta f(T)$ in the superconducting state in the L-orientation. As a
result, with due account of $\lambda_{ab}(T)$ (the inset to Fig.~2), we
derive from Eqs.~(3) and (5) $\lambda_c(0)$, which equals approximately
150~$\mu$m in sample~\#1. A similar procedure performed with sample~\#2
yields $\lambda_c(0)\approx 50$~$\mu$m, which is in agreement with our
measurements of overdoped BSCCO obtained using a different technique
\cite{Nic}. We also estimated $\lambda_c(0)$ on the base of absolute
measurements of the susceptibility $\chi'_c(0)$ from Eq.~(5), and we
obtained $\lambda_c(0)\approx 210$~$\mu$m for sample~\#1 and
$\lambda_c(0)\approx 70$~$\mu$m for sample~\#2. These results are in
reasonable agreement with our microwave measurements if we take into
consideration the fact that the accuracy of $\lambda_c$ measurements
is rather poor and the error can be up to 30\%.

In conclusion, we have used the ac-susceptibility and cavity perturbation
techniques in studying anisotropic high frequency properties
of BSCCO single crystals. We have observed almost linear dependences
$\Delta\lambda_c(T)\propto T$, which are in fair agreement with both
experimental \cite{Jac1,Shib2,Nic} and theoretical \cite{Klem}
results by other researchers. We have also investigated a new technique
for determination of $\lambda_c(0)$, which is a factor of three higher
in the optimally doped BSCCO sample than in the overdoped crystal. The
ratio between the slopes of curves of $\Delta\lambda_c(T)$ in the range
$T\ll T_c$ is the same. These facts could be put down to
dependences of $\lambda_c(0)$ and $\Delta\lambda_c(T)$ on the oxygen
content in these samples. At the same time, we cannot rule out influence
of defects in the samples on $\lambda_c(0)$, even though their quality
in the $ab$-plane is fairly high, according to our experiments.
In order to draw ultimate conclusions concerning the nature of the
transport properties along the $c$-axis in BSCCO single crystals, studies
of more samples with various oxygen contents are needed.

We would like to thank V.~F.~Gantmakher for helpful discussions. This
research was supported by grant No.~4985 of CNRS-RAS cooperation.
The work at ISSP was also supported by the Russian
Fund for Basic Research (grants 97-02-16836 and 98-02-16636) and
Scientific Council on Superconductivity (project 96060).


\vspace{-0.5cm}

\begin{references}
\vspace{-1.5cm}

\bibitem{Tru1} M.~R.~Trunin,~Physics--Uspekhi, {\bf 41}, 843 (1998);
J.~Superconductivity {\bf 11}, 381 (1998).

\bibitem{Scal} D.~J.~Scalapino, Phys. Rep. {\bf 250}, 329 (1995).

\bibitem{Shib1} T.~Shibauchi, H.~Kitano, K.~Uchinokura, A.~Maeda,
T.~Kimura, and K.~Kishio,
Phys. Rev. Lett. {\bf 72}, 2263 (1994).

\bibitem{Mao} J.~Mao, D.~H.~Wu, J.~L.~Peng, R.~L.~Greene, and S.~M.~Anlage,
Phys. Rev. B {\bf 51}, 3316 (1995).

\bibitem{Kit1} H.~Kitano, T.~Shibauchi, K.~Uchinokura, A.~Maeda,
H.~Asaoka, and H.~Takei, Phys. Rev. B {\bf 51}, 1401 (1995).

\bibitem{Bon1} D.~A.~Bonn, S.~Kamal, K.~Zhang, R.~Liang, and W.~N.~Hardy,
J. Phys. Chem. Solids {\bf 56}, 1941 (1995).

\bibitem{Jac1} T.~Jacobs, S.~Sridhar, Q.~Li, G.~D.~Gu, and N.~Koshizuka,
Phys. Rev. Lett. {\bf 75}, 4516 (1995).

\bibitem{Shib2} T.~Shibauchi, N.~Katase, T.~Tamegai, and K.~Uchinokura,
Physica C {\bf 264}, 227 (1996).

\bibitem{Srik}  H.~Srikanth, Z.~Zhai, S.~Sridhar, and A.~Erb,
J. Phys. Chem. Solids {\bf 59}, 2105 (1998).

\bibitem{Kit2} H.~Kitano, T.~Hanaguri, and A.~Maeda,
Phys. Rev. B {\bf 57}, 10946 (1998).

\bibitem{Hos} A.~Hosseini, S.~Kamal, D.A.~Bonn, R.~Liang, and W.~N.~Hardy,
Phys. Rev. Lett. {\bf 81}, 1298 (1998).

\bibitem{Nic} H.~Enriquez, N.~Bontemps, A.A.~Zhukov, D.V.~Shovkun,
M.R.~Trunin, A.~Buzdin, M.~Daumens, and T.~Tamegai,
submitted to Phys. Rev. B.

\bibitem{Tam}  S.~Ooi, T.~Shibauchi, and T.~Tamegai,
Physica C {\bf 302}, 339 (1998).

\bibitem{exp} We note that $\delta f(T)$ includes the frequency shift
due to the sample thermal expansion, which is essential for $T>0.7\,T_c$
in the T-orientation \protect\cite{Tru1}.

\bibitem{Lee}  S-F.~Lee, D.~C.~Morgan, R.~J.~Ormeno, D.~M.~Broun,
R.~A.~Doyle, and J.~R.~Waldram,
Phys. Rev. Lett. {\bf 77}, 735 (1996).

\bibitem{Gou} C.~E.~Gough and N.~J.~Exon, Phys. Rev. B {\bf 50}, 488 (1994).

\bibitem{Klem} R.~A.~Klemm and S.~H.~Liu, Phys. Rev. Lett.
{\bf 74}, 2343 (1995); T.~Xiang and J.~M.~Wheatley, {\it ibid.}
{\bf 76}, 134 (1996); R.~J.~Radtke, V.~N.~Kostur, and K.~Levin,
Phys. Rev. B {\bf 53}, R522 (1996).

\end{references}
\end{document}